\documentclass{svproc}

\usepackage{url}

\usepackage{graphicx}
\usepackage{xcolor}
\usepackage{listings}
\usepackage{float}

\begin{document}
\mainmatter              
\title{Optimized Real-Time Assembly in a RISC Simulator}
%
%
\author{Marwan~Shaban\inst{1} \and Adam~J.~Rocke\inst{2}}
\authorrunning{Marwan Shaban et al.} 
%
\tocauthor{Marwan Shaban and Adam J. Rocke}
\institute{Seminole State College of Florida, 100 Weldon Boulevard, Sanford, FL 32773, USA,\\
\email{shabanm@seminolestate.edu},
\and
Seminole State College of Florida, 100 Weldon Boulevard, Sanford, FL 32773, USA,\\
\email{rockea@seminolestate.edu}
}

\maketitle              

\begin{abstract}
Simulators for the RISC-V instruction set architecture (ISA) are useful for teaching assembly language and modern CPU architecture concepts. The {\em Assembly/Simulation Platform for Illustration of RISC-V in Education (ASPIRE)} is an integrated RISC-V assembler and simulator used to illustrate these concepts and evaluate algorithms to generate machine language code. In this article, ASPIRE is introduced, selected features of the simulator that interactively explain the RISC-V ISA as teaching aides are presented, then two assembly algorithms are evaluated. Both assembly algorithms run in real time as code is being edited in the simulator.  The optimized algorithm performs incremental assembly limited to only the portion of the program that is changed. Both algorithms are then evaluated based on overall run-time performance.
\end{abstract}
\section{Introduction}
Assembly language simulators are useful tools for illustrating both assembly language and CPU architecture topics.  They can also facilitate the evaluation of assembly algorithms to produce machine code for a given instruction set architecture (ISA).

In this paper, the {\em Assembly/Simulation Platform for Illustration of RISC-V in Education (ASPIRE)}\footnote{At the time of this writing, the source code is available at \url{https://github.com/ProfessorShaban/risc-v-simulator}.} is introduced and two assembly algorithms are evaluated. ASPIRE includes several teaching aides that interactively illustrate RISC-V assembly language (\cite{10.5555/3153875}, \cite{Waterman:EECS-2014-54}).  Features that are used to teach assembly language and CPU architecture concepts include graphical illustrations of instruction format, two's compliment signed integer format, and the IEEE 754 floating point number format.

The ASPIRE engine assembles the code in real time as it is being edited. Two algorithms to generate machine code are evaluated.  One algorithm assembles the entire program upon each character change in the code.  The second algorithm, an optimized version of the first, performs incremental updates by assembling only the modified portion and retaining the results for the unmodified portion of the program. The two versions are compared and runtime performance is evaluated.

\section{Related Work}

A handful of RISC-V assembler/simulator systems have been developed, most of them used in educational environments (see \cite{DBLP:journals/corr/abs-1908-09992}, \cite{jupiter}, \cite{rars}, \cite{9707149}, and \cite{waterman2011spike}). Incremental compilation has been used in integrated development environments (IDEs) for high-level languages (see \cite{10.1145/1006142.1006177}, \cite{10.1145/318593.318629}, \cite{10.1145/502874.502889}, and \cite{10.1145/502949.502887}), and to a limited extent for CISC assembly IDEs \cite{cuzzocrea}.

\section{Simulator Features}

ASPIRE is an integrated assembler and simulator. This allows the user to edit an assembly language program and observe changes made to register and memory values as the instructions are executed.  The base RISC-V integer instructions are supported, as well as the optional multiply-divide instruction extensions and floating-point instruction extensions. Other supported features include:

\begin{itemize}
  \item selected pseudoinstructions, such as j (jump) and mv (move),
  \item selected metainstructions, such as keyboard input, console output, and a stop instruction,
  \item data definition for strings, signed integers and floating point numbers, and
  \item labels to reference instructions and data.
\end{itemize}

\begin{figure}[tb]
\includegraphics[width=.9\linewidth]{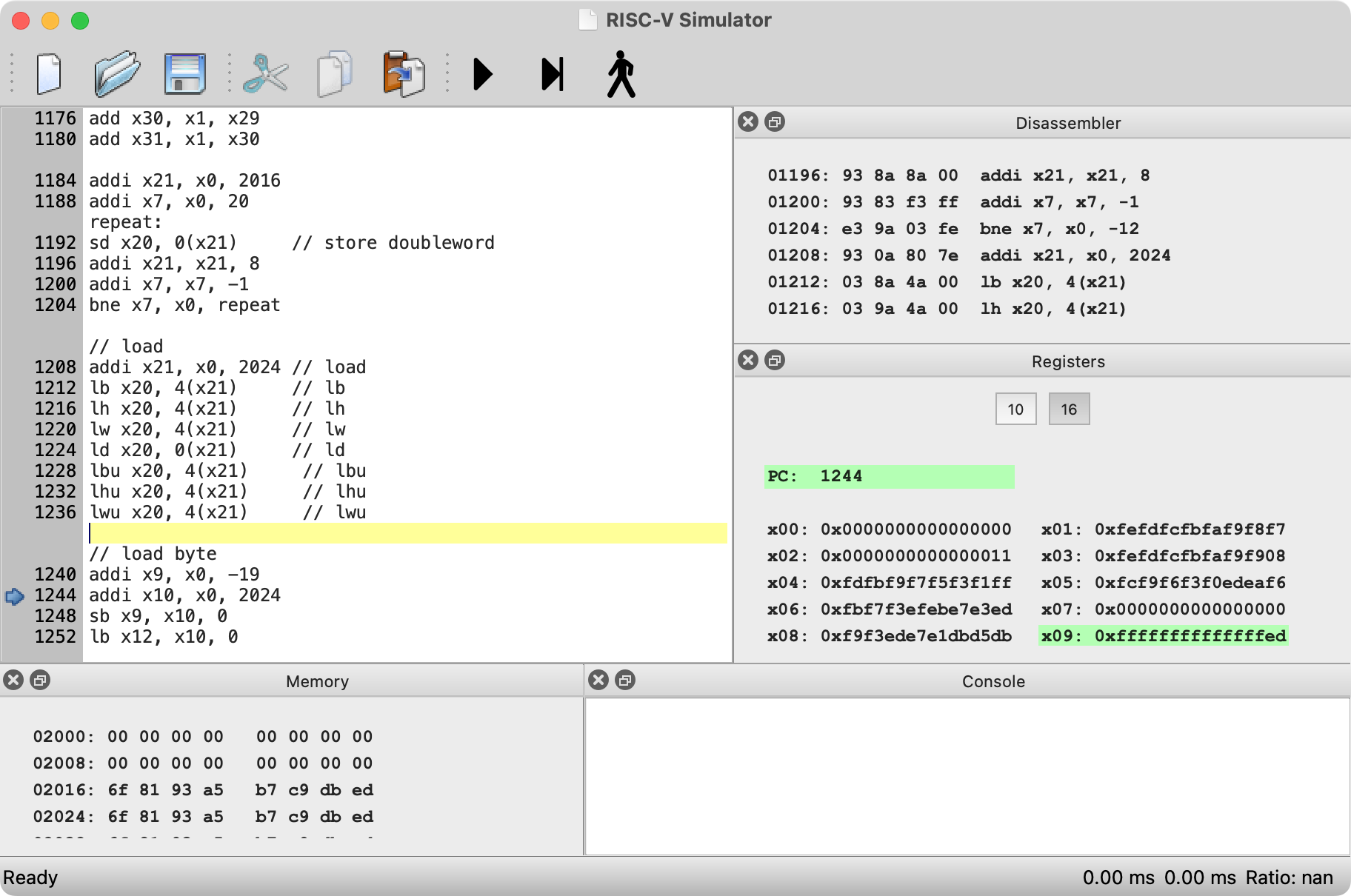}
\caption{Simulator overview}
\label{fig:01}
\end{figure}

Assembly is performed in real time as the program is modified.  This allows syntax errors to be highlighted as code is entered.  Execution of a program includes run, step, and animate commands. The run command executes the entire program from beginning to end, step will simulate execution of the next single instruction, and animate will execute instructions while displaying the execution results continuously. The main ASPIRE window is shown in Fig.~\ref{fig:01}.  Individual panes in the graphical environment show the contents of registers and memory and are updated in real time.  Registers and memory locations that have been modified by the last instruction to execute are highlighted. In addition, the disassembly pane shows the results of disassembling program memory.  This is particularly useful for explaining pseudoinstructions and relative addresses. 

\begin{figure}
\includegraphics[width=.9\linewidth]{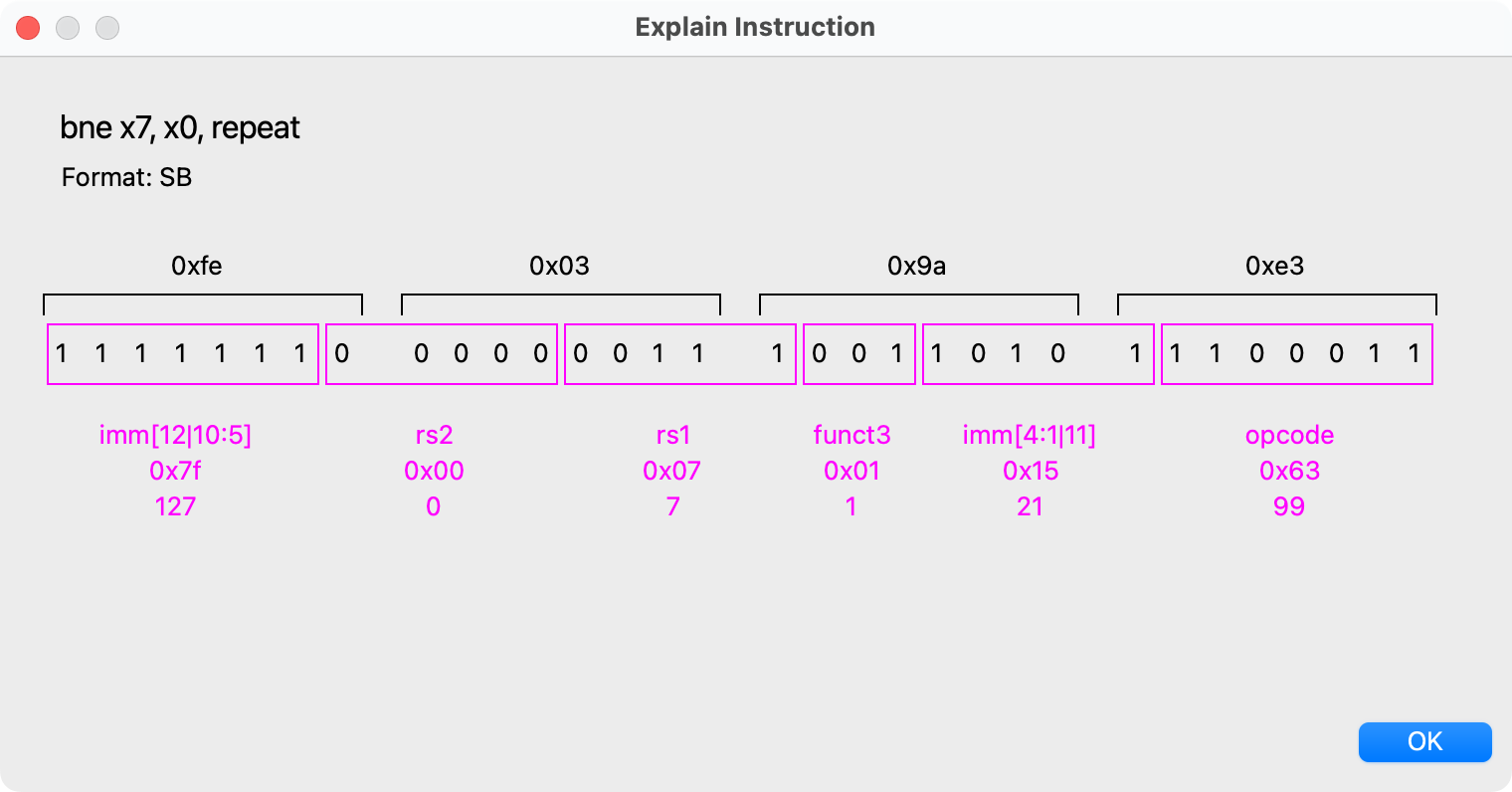}
\caption{The Explain Instruction dialog}
\label{fig:02}
\end{figure}

Context menu options allow the user to view the ``Explain Instruction'', ``Explain Floating Point Number'' and ``Explain Signed Integer'' dialog boxes.
The instruction encoding and format dialog box, shown in Fig.~\ref{fig:02}, displays the specific format and individual bit fields for a selected instruction.

\definecolor{commentsColor}{rgb}{0.497495, 0.497587, 0.497464}
\definecolor{keywordsColor}{rgb}{0.000000, 0.000000, 0.635294}
\definecolor{stringColor}{rgb}{0.558215, 0.000000, 0.135316}

\begin{figure}
\includegraphics[width=.9\linewidth]{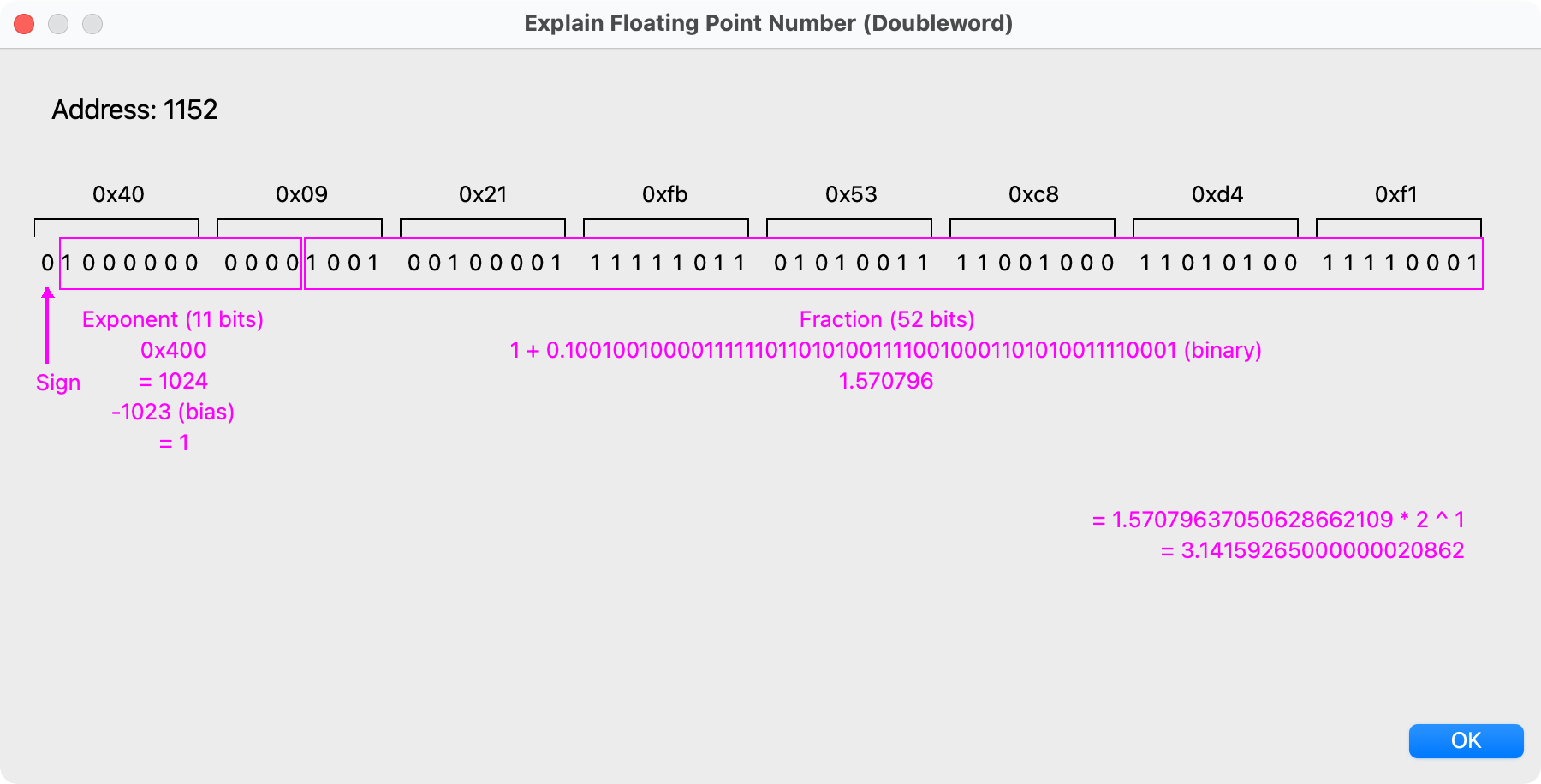}
\caption{The Explain Floating Point Number dialog}
\label{fig:03}
\end{figure}
\begin{figure}
\includegraphics[width=.9\linewidth]{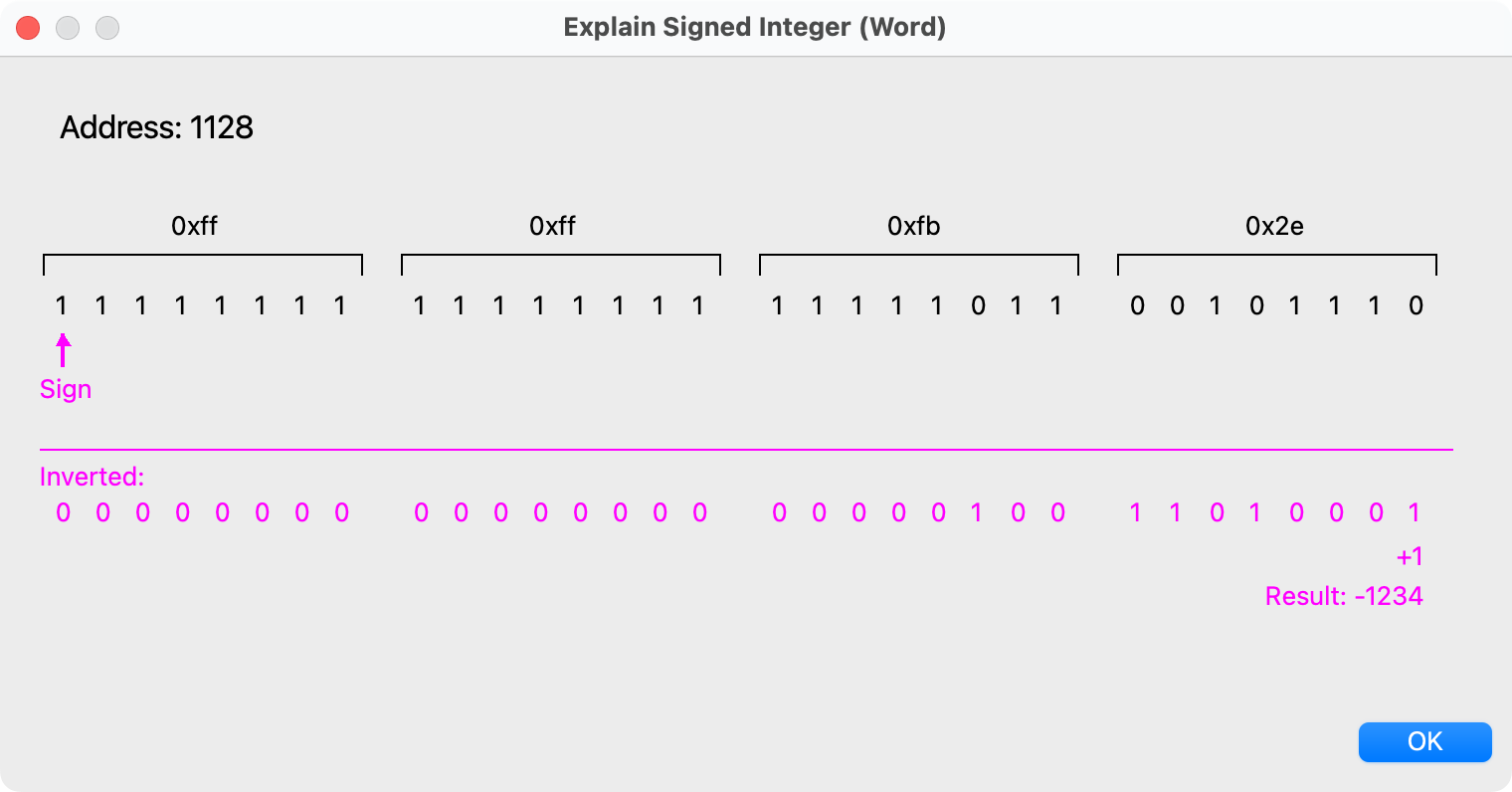}
\caption{The Explain Signed Integer dialog}
\label{fig:04}
\end{figure}

Encoding of floating point numbers is illustrated as shown in Fig.~\ref{fig:03} by defining the bit fields and other details for an IEEE 754 double-precision floating point number in memory. Fig.~\ref{fig:04} shows the Explain Signed Integer dialog box, which explains the format of a two's-compliment signed integer in memory.

\section{Assembler Optimization}

ASPIRE generates machine code directly from assembly so there is no intermediate object format. Thus, there is no need for a linker. The unoptimized algorithm will assemble the entire program upon each keystroke entered by the user. The optimized version performs incremental assembly by updating machine code in main memory as the assembly code is modified.  Assembling the unmodified portion of the program is not needed in the incremental version.  The consistent size of an assembly instruction in memory, four bytes or one 32-bit word, helps to simplify the incremental assembly algorithm.

As the program is being modified, the engine performs incremental assembly for most keystrokes.  Full assembly is performed upon cut, paste, delete, and when editing data statements or labels. Thus, most of the user's keystrokes only require incremental assembly resulting in the speedup for the optimized version.

Two of the important data structures used by the assembly engine are the line table and the symbol table.  The line table maps individual assembly instructions to binary code addresses. Each row in the line table corresponds to a single instruction in the source program, and contains an instance of the C structure shown in Fig.~\ref{fig:ais}.  The table is updated whenever a source line is modified.  Each entry in the line table records:

\begin{itemize}
	\item the content of the source line,
	\item the line number in the source file,
	\item the corresponding binary address,
	\item a valid flag to indicate a well formed assembly instruction, and
	\item additional details of the instruction such as the opcode and immediate fields.
\end{itemize}

The symbol table is used to track labels used in the source program and is updated in response to a change in a label or a reference to a label. Each entry in the symbol table records:

\begin{itemize}
	\item the label,
	\item the line number where the label is declared,
	\item the memory address that the label refers to, and
	\item a linked list of references to the label, each containing the reference's line number in the source file and the binary address at which the reference is located.
\end{itemize}

When a new instruction is inserted, the ASPIRE assembly engine will:

\label{insertingnew}
\begin{itemize}
	\item assemble the instruction and update the line table (this includes updating the symbol table if the instruction has a symbol reference),
	\item insert the corresponding machine code in memory moving all subsequent bytes down by one word,
	\item update the addresses and line numbers in the line table of all lines after the inserted instruction,
	\item update the addresses and line numbers in the symbol table of all labels and label references after the inserted instruction, and
	\item fix relative addresses belonging to label references that cross the new line, either as a forward or backward reference.
\end{itemize}

\begin{figure}
\begin{lstlisting}[basicstyle=\footnotesize, language=C, keywordstyle=\color{keywordsColor}\bfseries]
typedef struct {
    int source_line_number;
    int error;
    char *error_message;
    unsigned long long address;
    int length;
    char source_line[SOURCE_LINE_MAX+1];
    char *mnemonic;
    char format;
    unsigned int instruction;
    unsigned int opcode;
    unsigned int funct7;
    unsigned int funct3;
    unsigned int rs1, rs2, rd;
    int imm31;
    int imm11;
} assembly_instruction;
\end{lstlisting}
\caption{C structure representing an assembly instruction}
\label{fig:ais}
\end{figure}


\section{Runtime Complexity Analysis}

The environment can be configured to use either incremental assembly or full assembly upon each keystroke entered by the user.  The default algorithm is incremental assembly.

\subsection{Full Assembly Mode}
\label{fullmode}

In full assembly mode, the entire program is assembled upon each keystroke entered by the user. Assembling each line requires translation of the assembly instruction to machine code.  If the instructions contains a symbol reference, the symbol table has to be updated to add the symbol reference. With regard to execution time:

\begin{itemize}
	\item assembling a single line into the equivalent machine code is a constant-time operation,
	\item finding the symbol in the symbol table is linear in the size of the symbol table, and
	\item adding the symbol reference is a constant-time operation since adding an item to a linked list takes constant time.
\end{itemize}

Locating symbols in the symbol table can be optimized by sorting the symbol table by symbol name.  This may in turn result in faster assembly if there are a large number of symbol references.

Let $n$ be the number of lines, and $m$ be the number of labels. The symbol table will have $m$ entries. Given the above runtimes, the time to assemble the program is $O(n~m)$.

\subsection{Incremental Assembly Mode}

In incremental assembly mode, a keystroke entered by the user on a particular line of source code will cause only that line to be re-assembled. But, in incremental assembly mode, the engine will revert to full assembly upon certain keystrokes including:

\begin{itemize}
	\item deleting, cutting, or pasting, which may change multiple lines,
	\item entering a colon, meaning the line becomes a label declaration, and
	\item changes to data statements, which are lines declaring constant data used by the program.
\end{itemize}

The majority of keystrokes do not fall in the above categories and will result in incremental assembly.

When an empty line is inserted into the source code, an entry is inserted into the line table, but the machine code is unchnaged.

Each line in the source program has a corresponding word in the machine code, except for empty lines, labels, data statements and comments. Inserting a new assembly instruction causes a word to be inserted into the machine code, even if it is not a valid instruction. When the assembly instruction is invalid, its corresponding word in the machine code is a placeholder as opposed to a valid RISC-V instruction.

When a line in the assembly program is changed from being empty to having one character, it is assembled and four bytes are inserted into the machine code. As mentioned above, if it is not a well-formed instruction, these four bytes are simply a placeholder.  Thus, the process of actually inserting a word into the machine code happens only when a new, non-empty line appears. In addition to inserting four bytes in the machine code, relative addresses in the machine code representing label references are adjusted for each symbol in the symbol table.  This is necessary for each label where the inserted line occurs between the symbol declaration and a reference to it.  Due to locality of reference, only a small number of symbol references cross a given line, but each symbol reference in the symbol table must be examined. It's possible that a novel symbol table structure can reduce this cost by speeding up the lookup of symbol references crossing a given memory address, increasing performance when the symbol table is large.

Processing a change to an existing assembly instruction in incremental assembly mode is similar to the process outlined in Sect.~\ref{fullmode} for full assembly mode, taking $O(m)$ runtime in the worst case.

When a new instruction is inserted, the full version of the assembler will operate in $O(n~m)$ time, as described in Sect.~\ref{fullmode}, where $n$ is the number of lines and $m$ is the number of symbols. In incremental assembly mode, inserting an instruction causes the assembler to follow the process outlined earlier, which is reproduced here with runtime analysis.

\begin{enumerate}
	\item Assemble the instruction and update the line table. This includes updating the symbol table if the instruction has a symbol reference. The runtime is constant if the instruction doesn't have a label reference, and $O(m)$ if the instruction has a label reference, where $m$ is the number of symbols in the symbol table.
	\item Insert the corresponding machine code in memory moving all subsequent bytes down by one word. This runs in $O(n)$ time, where $n$ is the number of lines in the program.
	\item Update the addresses and line numbers in the line table for all lines after the inserted instruction. This also runs in $O(n)$ time, where $n$ is the number of lines in the program.
	\item Update the addresses and line numbers in the symbol table for all labels and label references after the inserted instruction. This runs in $O(m~r)$ time, where $m$ is the number of symbols in the symbol table and $r$ is the average length of the linked lists of references. Note that $O(m~r)$ can also be expressed as $O(m) + O(r_t)$ where $r_t$ is the total number of references.
	\item Fix relative addresses belonging to label references that cross the new line, either as a forward or backward reference. This also runs in $O(m~r)$ time, where $m$ is the number of symbols in the symbol table and $r$ is the average length of the linked lists of references.
\end{enumerate}

Thus, the overall runtime in the worst case of inserting a new assembly instruction in the incremental assembler is $O(n) + O(m~r)$.

Note that symbol references are only used to adjust relative addresses, and adjusting an address is done by re-assembling the line at which the reference occurs. So, when a line having a reference is changed such that the reference is removed, it’s okay to leave the old reference in the symbol table rather than incur the cost of scanning for unused symbol references.

The full and incremental versions, when inserting a new assembly instruction, run at worst in $O(n~m)$ and $O(n) + O(m~r)$ time, respectively. However, the two versions have drastically different constant coefficients ($c_{1}$ in equation \ref{eq:eq1}), where

\begin{equation}
\label{eq:eq1}
O(n) = c_1 * n + c_2
\end{equation}

In the case of full assembly, $c_{1}$ includes full assembly of each source line upon each keystroke. In the case of incremental assembly, $c_{1}$ includes shifting bytes, and adjusting addresses in the line table as previously mentioned, only upon the first keystroke that transitions the line from empty to nonempty.

Table~\ref{complexitycomparison} summarizes the runtime complexity of each version when processing a single keystroke, for each of the three main cases (changing an instruction, adding an instruction, and all other changes, e.g., deleting an instruction).

\begin{table}[ht]
\caption{Assembly runtime for a single keystroke}
\begin{tabular}{rrrrr}
Type of change & \space & Full Assembler & & Incremental Assembler\\
Changing an instruction & & $O(n~m)$ & & $O(m)$ \\
Adding an instruction & & $O(n~m)$ & & $O(n) + O(m~r)$ \\
All other changes & & $O(n~m)$ & & $O(n~m)$ \\
\end{tabular}
\label{complexitycomparison}
\end{table}

Results comparing the execution time for the two assembler versions when adding an instruction are shown in Sect.~\ref{timingsection} below.


\section{Timing Comparison}
\label{timingsection}

\begin{figure}
\includegraphics[width=0.6\linewidth]{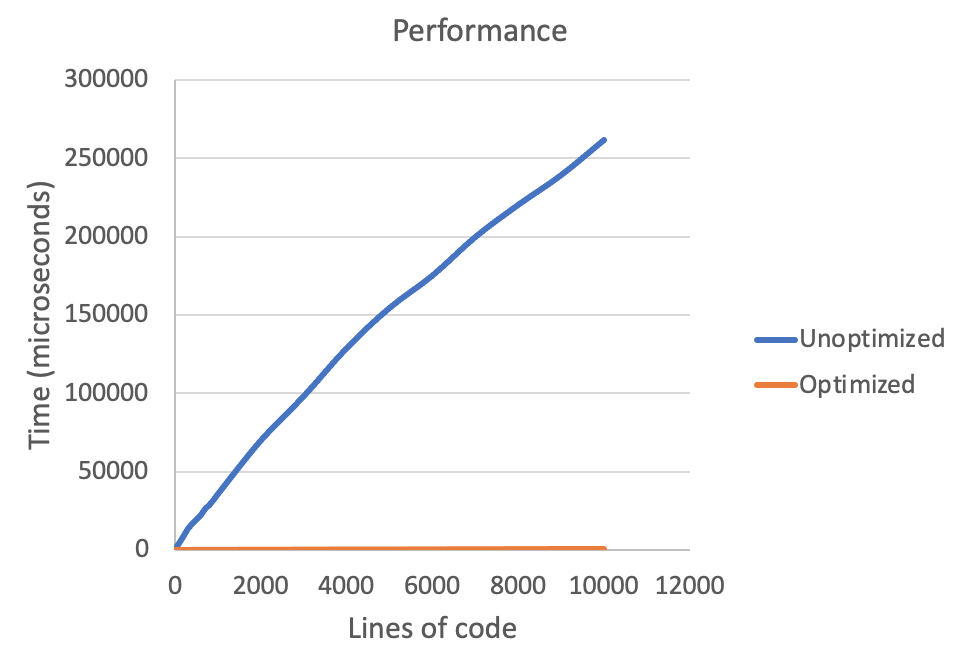}
\caption{Performance comparison}
\label{fig:05}
\end{figure}

Multiple tests were performed on a MacBook Pro with a 2.2 GHz Intel Core i7 processor using different input sizes.  Each test consisted of inserting the single instruction ``addi x1, x2, -121'' in the middle of an existing source file. The timing results shown in Table \ref{timingtable} and Fig.~\ref{fig:05} illustrate the time to process a change in both the full and incremental versions, and shows that the incremental assembly algorithm is far more efficient than the full assembly algorithm. Note that a series of tests that only involve changing an instruction that does not have a label reference would have constant runtime in the incremental assembler, with even faster performance versus the full assembler.

\begin{table}[ht]
\caption{Timing Results}
\tiny
\scriptsize
\begin{tabular}{rrr}
Lines & Unoptimized & Optimized\\
 & (microseconds) & (microseconds)\\
1 & 110 & 142 \\
10 & 654 & 133 \\
100 & 4544 & 134 \\
1000 & 35712 & 165 \\
2000 & 69889 & 205 \\
3000 & 98084 & 228 \\
4000 & 128727 & 284 \\
5000 & 154417 & 280 \\
6000 & 175045 & 325 \\
7000 & 199652 & 363 \\
8000 & 220164 & 385 \\
9000 & 239106 & 429 \\
10000 & 261555 & 424 \\
\end{tabular}
\normalsize
\label{timingtable}
\end{table}

\section{Conclusions}

The results obtained via simulation show that an integrated assembler/simulator can perform efficient incremental assembly of assembly language source files in real time.  This promotes interaction with the simulator environment and helps facilitate teaching assembly language and modern CPU architecture concepts.

%
%
\bibliographystyle{acm}
\bibliography{refs}

\end{document}